\documentclass[letterpaper,10pt]{paper}
\usepackage[utf8]{inputenc}
\usepackage{amsmath}
\usepackage{booktabs}
\usepackage{stmaryrd}
\usepackage{amssymb}
\usepackage{graphicx}
\usepackage{amsthm}
\usepackage{fancyvrb}
\usepackage{calc}
\usepackage{tikz}
\usepackage{url}

\title{Correlation and Substitution in SPARQL}
\author{Daniel Hernández \and Claudio Gutierrez \and Renzo Angles}

\newtheorem{lemma}{Lemma}

\theoremstyle{definition}
\newtheorem{definition}{Definition}

\newcommand{\semantics}[1]{\ensuremath{\llbracket #1 \rrbracket}}

\newcommand{\sigla}[1]{{\sc #1}}
\newcommand{\dom}{\operatorname{dom}}
\newcommand{\norm}{\operatorname{norm}}
\newcommand{\bind}{\operatorname{bind}}

\newcounter{query}
\newcommand{\query}[1]{\refstepcounter{query}\label{#1}}
\DefineVerbatimEnvironment%
  {Query}{Verbatim}
  {frame=lines,framesep=2mm,numbers=left,samepage=true,fontsize=\small}

\begin{document}

\maketitle

\noindent
\textsf{Version \today}

\begin{abstract}
  In the current \sigla{sparql} specification the notion of
  correlation and substitution are not well defined. This problem
  triggers several ambiguities in the semantics.  In fact,
  implementations as Fuseki, Blazegraph, Virtuoso and rdf4j assume
  different semantics.

  \medskip

  In this technical report, we provide a semantics of correlation and
  substitution following the classic philosophy of substitution and
  correlation in logic, programming languages and \sigla{sql}. We
  think this proposal not only gives a solution to the current ambiguities and
  problems, but helps to set a safe formal base to further extensions
  of the language.

  \medskip

  This work is part of an ongoing work of Daniel Hernandez.  These
  anomalies in the W3C Specification of SPARQL 1.1 were detected early
  and reported no later than 2014, when two erratas were registered
  (cf. \url{https://www.w3.org/2013/sparql-errata#errata-query-8} and
  \url{https://www.w3.org/2013/sparql-errata#errata-query-10}).
\end{abstract}

\clearpage

\section{Introduction}

The first version of this technical report served as a starting point
to restart the discussion about the substitution and correlation in
SPARQL. This issue was discussed in several threads on the W3C
{public-sparql-dev} mailing list (see messages of Jun, 2016 in the
mailing list
archives\footnote{\url{https://lists.w3.org/Archives/Public/public-sparql-dev/2016AprJun/}})
and a W3C Community Group\footnote{\url{https://www.w3.org/community/sparql-exists/}}
was created to discuss and address problems with the specification
of the \texttt{EXISTS} clause in SPARQL.

 This new version of the report fixes errors of the previous one
and includes a formalization of two alternative semantics that are 
currently implemented: The first by Blazegraph and Fuseki,  a semantics 
where substitution is never applied because variables that are not projected 
to resulting solutions are not visible from outside. The second,
 by Virtuoso and rdf4j, where every variable that is not
projected to resulting solution are visible from outside, so they can
be substituted.
 Within the same formal framework, we show the semantics presented in 
the previous technical report, where some
variables are visible and other are not.

  The main idea of the formal framewok work as follows.
 Given a graph pattern \texttt{P} and a solution mapping $\mu$, 
the Standard Spec. of SPARQL introduces the notion of 
substitute$(\mathtt{P},\mu)$, that is used to evaluate nested
patterns. However, as we show in this report, this function
substitute is not well defined and is contradictory with other
parts of the specification. In this tech report, we define 
a similar function, which we call bind (to avoid clash names),
that solves the problems found. It basically normalizes the
pattern $\mathtt{P}$ before applying the mapping $\mu$, 
giving a new structure $\norm(P)$ that essentially renames variables
so that each one plays the same role in every occurrence.

\paragraph{Structure of this technical report} Section
\ref{sec:evaluation-of-correlated-variables} presents an example of
correlation using substitution to exemplify ambiguities of the current
specification and differences of implementations. Section
\ref{sec:problems} describes the problem with current notion of
substitution. In Section \ref{sec:semantics} we propose three
alternative ways to define $\bind(\mathtt{P},\mu)$ based in
alternative definitions for $\norm(\mathtt{P})$. Section
\ref{sec:blank-nodes} discuss how the proposed semantics are safe
regarding with the use of blank nodes. Finally, in Section
\ref{sec:examples} we present several examples that illustrated how
correlation is evaluated in each semantics and how implementations
match them.

\section{Evaluation of correlated variables}
\label{sec:evaluation-of-correlated-variables}

Consider the following simple \sigla{sparql} query that 
selects people of country $j$ that have children, and consider as data
the \sigla{rdf} graph depicted in Figure \ref{fig:rdf-graph} below.

\query{query:sparql-parents-simple}
\begin{Query}[label=\textrm{Listing \thequery}]
SELECT ?parent
WHERE { ?parent :country :j
        FILTER ( EXISTS { SELECT ?child
                          WHERE { ?child :parent ?parent }})}
\end{Query}

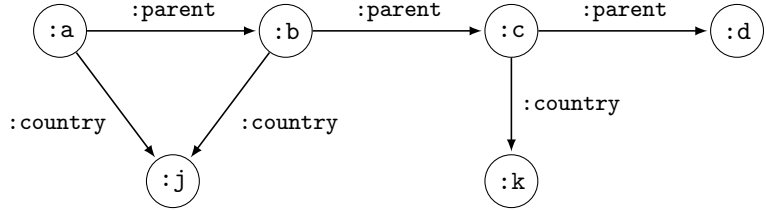
\begin{figure}[h]
\begin{center}
\begin{tikzpicture}
\begin{scope}[auto, every node/.style={draw,circle,minimum size=2em,inner sep=1},node distance=2cm]
\def\margin {8}
 \node[draw,circle](a) at (0,0) {\texttt{:a}};
 \node[draw,circle](j) at (1.5,-2) {\texttt{:j}};
 \node[draw,circle](b) at (3,0) {\texttt{:b}};
 \node[draw,circle](c) at (6,0) {\texttt{:c}};
 \node[draw,circle](k) at (6,-2) {\texttt{:k}};
 \node[draw,circle](d) at (9,0) {\texttt{:d}};
\end{scope}

\begin{scope}[->,>=latex,semithick,shorten >=1pt, auto]
\draw(a) to node {{\small\texttt{:parent}}} (b);
\draw(b) to node {{\small\texttt{:parent}}} (c);
\draw(c) to node {{\small\texttt{:parent}}} (d);
\draw(a) to node[swap] {{\small\texttt{:country}}} (j);
\draw(b) to node {{\small\texttt{:country}}} (j);
\draw(c) to node {{\small\texttt{:country}}} (k);
\end{scope}
\end{tikzpicture}
\end{center}
\caption{RDF graph.}
\label{fig:rdf-graph}
\end{figure}

The engines Fuseki and Blazegraph give as solution two mappings:
$\mu_a$ and $\mu_b$, where $\mu_a$ is
$\{\text{\tt?parent}\mapsto\text{\tt:a}\}$ and $\mu_b$ is
$\{\text{\tt?parent}\mapsto\text{\tt:b}\}$.  Virtuoso and rdf4j
(formerly Sesame), on the other hand, give a different result, only
$\mu_b$.  Why these differences?

\medskip

 What does the W3C Spec. tell?
The query  has the form
\texttt{SELECT ?parent WHERE \string{P FILTER (EXISTS \string{Q\string})\string}},
where we will call \texttt{P} and \texttt{Q}  respectively the outer and the inner graph
patterns. 
How to evaluate this query?
The W3C Spec. gives these two definitions that are relevant for this:

\begin{definition}[Standard substitution, W3C Spec., \S18.6]
\label{def:standard-substitution}
Let $\mu$ be a solution mapping an \texttt{P} be a graph
pattern. Then, $\mathrm{substitute}(\mathtt{P},\mu)$ is the graph
pattern formed by replacing, for each $x$ in $\mathrm{dom}(\mu)$,
every occurrence of a variable $x$ in \texttt{P} by $\mu(x)$.
\end{definition}

\begin{definition}[Evaluation of Exists, W3C Spec. \S18.6]
\label{def:exists}
Let $\mu$ be the current solution mapping for a filter and \texttt{P}
a graph pattern: The value $\mathrm{exists}(\mathtt{P})$, given $D(G)$,
is true if and only if
$\mathrm{eval}(D(G),\mathrm{substitute}(\mathtt{P}, \mu))$ is a
non-empty sequence.
\end{definition}

In definition \ref{def:exists} above, the argument $D(G)$ denotes that
the evaluation is done against the dataset $D$ using the graph
$G$. For the sake of the readability, in that follows we use the
notation $\semantics{\mathtt{P}}_D$ instead of $\mathrm{eval}(D(G),
\mathtt{P})$.

\medskip\noindent
The result of the \texttt{P FILTER (EXISTS \string{Q\string})} clause, according to the W3C Spec.,
should be the set $\Omega$ such that:\footnote{ We will avoid the
  multiplicities in this report because the problems we report are
  independent of having set or multiset semantics.  }
\begin{align*}
\Omega &=
\{
\mu \in \semantics{\mathtt{P}}_D
\mid
\semantics{\mathrm{substitute}(\mathtt{Q},\mu)}_D \text{ is not empty}
\}.
\end{align*}

\medskip

So, it seems that Virtuoso and rdf4j follow the standard here: First,
evaluate the pattern \texttt{P}, which give mappings $\mu_a$ and
$\mu_b$, and for each of them, perform the replacement in \texttt{Q}.
As $\semantics{\mathrm{substitute}(\mathtt{Q},\mu_a)}_D = \varnothing$ and
$\semantics{\mathrm{substitute}(\mathtt{Q},\mu_b)} \neq\varnothing$, the final solution is
the mapping $\mu_b$.

In defense of Fuseki and Blazegraph, let us say that the W3C Spec. says
in other place (\S12. Subqueries): ``Note that only variables projected
out of the subquery will be visible, or in scope, to the outer
query.'' That is, the variable \texttt{:parent} inside the
\texttt{WHERE} clause is not visible from outside, and thus, 
\texttt{Q} cannot be changed by any mapping $\mu$
(in the sense of Defn. \ref{def:standard-substitution}). Thus the \verb+FILTER (EXISTS {Q})+ is true, and thus
the two mappings $\mu_a$ and $\mu_b$ qualify as final solutions.

\medskip

The situation becomes even more involved when one considers another way 
of writing the previous query. 
Consider now the query in listing \ref{query:sparql-parents}.

\query{query:sparql-parents}
\begin{Query}[label=\textrm{Listing \thequery}]
SELECT ?parent
WHERE { ?parent :country :j
        FILTER ( EXISTS { SELECT ?child
                          WHERE { ?child :parent ?chparent
                                  FILTER (?chparent = ?parent) }})}
\end{Query}

The listings \ref{query:sparql-inner-1} and \ref{query:sparql-inner-2}
present the queries resulting of applying the substitution on the
inner graph pattern with the mappings $\mu_a$ and $\mu_b$, respectively.

\bigskip

\begin{flushleft}
\begin{minipage}{0.45\textwidth}
\query{query:sparql-inner-1}
\begin{Query}[label={\rm Listing \thequery}]
SELECT ?child
WHERE {
  ?child :parent ?chparent
  FILTER (?chparent = :a)
}
\end{Query}
\end{minipage}%
\hfill
\begin{minipage}{0.45\textwidth}
\query{query:sparql-inner-2}
\begin{Query}[label={\rm Listing \thequery},numbers=none]
SELECT ?child
WHERE {
  ?child :parent ?chparent
  FILTER (?chparent = :b)
}
\end{Query}
\end{minipage}
\end{flushleft}

\medskip

Only the second listing has solutions, so the evaluation of whole query returns
$\{\mu_b\}$.  Actually, in Virtuoso this query outputs the expected
result, that is, $\{\mu_b\}$.  On the contrary, in Fuseki and
Blazegraph this query outputs no solutions.  These systems are following
another part of the W3C Spec. (12. Subqueries): ``Due to the
bottom-up nature of SPARQL query evaluation, the subqueries are
evaluated logically first, and the results are projected up to the
outer query.''  Thus, they probably consider that the inner query
returns error because there is a non-bound variable \texttt{?parent}.

\paragraph{Engines} 
Examples presented in this report where tested in Fuseki
2.4.0\footnote{\url{https://jena.apache.org}}, Blazegraph Community
Edition 2.1.0\footnote{\url{https://www.blazegraph.com}}, Virtuoso
Open Source Edition
7\footnote{\url{https://github.com/openlink/virtuoso-opensource/tree/stable/7}}
and rdf4j 2.0 Milestone Builds\footnote{\url{http://rdf4j.org}}.

\section{Problems with the current notion of substitution}
\label{sec:problems}

The previous examples show that problems arise when is not clear if
occurrences of a variable are correlated\footnote{We use
  ``correlated variables'' and
  ``correlation'' to indicate the occurrence of a variable $x$ in and
  expression $E$ whose value depends on the value of the occurrence of
  same variable $x$ in an expression containing $E$. The paradigmatic
  occurrence of correlation in SPARQL is the expression
  \texttt{Q EXIST FILTER (P)}.}
and, in particular, if substitution has to be applied in a
variable. In this section we will show that one of the main problems
with nested queries in \sigla{sparql} is that the notion of
{\em substitution} is not well defined. We will present a solution to
this issue, which in turns helps to fix the whole semantics of
nesting.

First, consider the simple graph pattern \texttt{Q}:
\[
\text{\tt SELECT ?x WHERE \string{:a :p ?x\string}}
\]
and let $\mu$ be the solution
mapping $\{\text{\tt?x}\mapsto 1\}$. Then, $\mathrm{substitute}(\mathtt{Q},\mu)$, understood
literally from the standard, means replacing every occurrence of
\texttt{?x} with 1, that gives:
\[
\text{\tt SELECT 1 WHERE \string{:a :p 1\string}}
\]
Thus, the substitution method of the W3C Spec. is not
well defined because it breaks the grammar of the \texttt{SELECT} clause.\footnote{
This and more subtle problems that a naive notion of substitution brings
are well know in logic and algebra long ago. 
For example, a variable $x$ cannot be substituted by a constant
 in all its  occurrences in the first order formula $\forall x \,
 p(x)$ or in an expression like $\sum_{x \in A}(x+a)$.
}.

 The notion of substitution was already present in SPARQL 1.0 in
 patterns of the form \texttt{P FILTER (C)}.
In SPARQL 1.0 \texttt{C} is a Boolean clause, and here the
substitution works fine
because every occurrence of a variable could be replaced without
breaking the grammar. The only case that required a special treatment was
the function \texttt{bound(?x)} where \texttt{?x} was not substituted, but
checked if it was in the domain of the current solution.

On the contrary, in SPARQL 1.1, the clauses \texttt{EXISTS
  \string{Q\string}} and \texttt{NOT EXISTS \string{Q\string}} are
filter constraints, thus allow nesting a graph pattern \texttt{Q}
instead of the Boolean clause in a filter.  And a graph pattern may
contain variables with occurrences that are not replaceable (as we
previously discussed) and variables with occurrences that are not
``visible from outside'' \texttt{Q}. Thus, the naive substitution,
consisting on just replacing all occurrences of a variable, cannot be
directly applied in SPARQL 1.1 as was in SPARQL 1.0.

\section{Semantics of nested expressions with correlated variables}
\label{sec:semantics}

In this section we propose three alternatives for the function
$\bind$, that is defined as alternative to the function
$\mathrm{substitute}$ for evaluating nested queries with correlated
variables, two of which represent approaches existing in current
implementations and the other the approach that was proposed in the
previous version of this report.

\subsection{The domain of a graph pattern}

Given a graph pattern \texttt{P}, we denote as
$\mathrm{var}(\mathtt{P})$ to the set of variables that occur in
\texttt{P}.

An interesting subset of $\mathrm{var}(\mathtt{P})$ is the one that
includes the variables that occur in the solutions of \texttt{P}, that
we call the domain of \texttt{P} and denote as
$\mathrm{dom}(\mathtt{P})$, that is formally defined as
follows\footnote{Recall that $\dom(\mu)$ is the domain of variables of
  $\mu$ (those where $\mu$ is defined) when $\mu$ is considered as a
  partial function over the set of all variables of the universe.}:
\[
 \mathrm{dom}(\mathtt{P}) = 
   \{ x \in \dom(\mu)\; |   \text{ exists dataset $D$ with } \mu \in \semantics{\mathtt{P}}_D. \}
\]

This definition of dataset cannot be used directly to compute the
domain of a graph pattern, because requires the verification in all
possible datasets. The following lemma shows that it is possible to
give a method to compute the domain of a graph pattern using only its
syntax.

\begin{lemma}[In-domain variables]
  Given a graph pattern \texttt{P} and a variable \texttt{?x}
  occurring in \texttt{P}, then
  $\text{\tt?x}\in\mathrm{dom}(\mathtt{P})$ if and only if:
\begin{enumerate}
\item \texttt{P} is a basic graph pattern and \texttt{?x} occurs in
  \texttt{P}.
\item If \texttt{P} is \texttt{Q$\circ$R} where $\circ$ is `\texttt{.}',
  \texttt{UNION} or \texttt{OPTIONAL} and
  $\text{\tt?x}\in\mathrm{dom}(\mathtt{Q})\cup\mathrm{dom}(\mathtt{R})$.
\item If \texttt{P} is \texttt{Q MINUS R} where $\circ$ is `\texttt{.}',
  and $\text{\tt?x}\in\mathrm{dom}(\mathtt{Q})$.
\item If \texttt{P} is \texttt{GRAPH ?x \string{Q\string}}.
\item If \texttt{P} is \texttt{GRAPH u \string{Q\string}} and
  $\text{\tt?x}\in\mathrm{dom}(\mathtt{Q})$.
\item If \texttt{P} is \texttt{VALUES (X) \string{B\string}} and
  \texttt{?x} occurs in the list of variables \texttt{X}.
\item If \texttt{P} is \texttt{Q BIND (E AS ?x)}.
\item If \texttt{P} is \texttt{Q BIND (E AS ?y)}
  and $\text{\tt?x}\in\mathrm{dom}(\mathtt{Q})$.
\item If \texttt{P} is \texttt{Q FILTER (C)}
  and $\text{\tt?x}\in\mathrm{dom}(\mathtt{Q})$.
\item If \texttt{P} is \texttt{SERVICE u \{Q\}}
  and $\text{\tt?x}\in\mathrm{dom}(\mathtt{Q})$.
\item If \texttt{P} is \texttt{SELECT X WHERE \{Q\}} if \texttt{X} is
  a list such that one of its elements is \texttt{?x} or has the form
  \texttt{(E AS ?x)}.
\end{enumerate}
\end{lemma}

Note that in the SPARQL specification variables that are in the domain
of graph pattern are called {\em in-scope} and also defined using the
syntax (see 18.2.1 in the W3C Spec.). In this report we call
them {\em in-domain} to stress the idea that they define the domain of the
output.

\subsection{Syntax and variables roles}

A variable in the syntax can play several roles. 
Two relevant ones are the role of representing the the output of a computation
(output role) and the one representing the output of a previous computation (input
role). For example, in the expression \texttt{let x be f(y) in \{  g(x,y) \}} in a functional
language, the variable \texttt{x} is playing the
output role in the outermost occurrence and the input role in the
innermost occurrence.  On the other hand, \texttt{y} plays the input role in both
occurrences.

These roles are crucial to understand how the substitution of variables
by values work. Indeed, when a variable is in the input role,  we can
substitute it without breaking the syntax of the language. On the
contrary, a variable in the output role cannot be substituted, because
values cannot be used to name results of computations.

The question that arises is if we can distinguish the role of a
variable occurrence in SPARQL. To answer this question, let us to
consider the following types of syntactic constructs
in which variables occur:
\begin{description}
\item[Expression.] In a comparison (e.g., \texttt{?x < 2}),
  an scalar operation (e.g., \texttt{1+x}) or an scalar function
  (e.g., \texttt{substr(?x, 4)}).
\item[Pattern.] In a basic graph pattern (e.g., \texttt{?x :p ?y}).
\item[Naming.] In any place that only variables are allowed (e.g.,
  \texttt{E AS ?x}), except when they occur in the
  \texttt{bound($\cdot$)} function (e.g., \texttt{bound(?x)}), that is
  an special case.
\end{description}
The ocurrence of variables in the three types of 
constructs are associated with the output or input roles as shown in
Table \ref{table:role-occurrences}.

\begin{table}[h!]
\begin{center}
\begin{tabular}{lcc}
\toprule
& Input & Output \\
\midrule
Expression & $\times$ &\\
Pattern & $\times$ & $\times$ \\
Naming && $\times$ \\
\bottomrule
\end{tabular}
\end{center}
\caption{Possible variable roles in types of syntactic constructs.}
\label{table:role-occurrences}
\end{table}

In occurrences in expressions, it is clear that the variable
can be substituted, because it refers to a value to be used inside the
expression. Similarly, in naming occurrences, it is clear that the
variable cannot be substituted, because breaks the grammar. Moreover,
the variable will be used to refer the value in a future computation
so its name cannot be forgotten nor changed.

In the case of occurrences in patterns the variable could have both
roles. Indeed, we can substitute the variable with a value without
breaking the semantics, so the variable is playing the input
role. On the other hand, if the variable is not substituted, then it
will bind a value from the data that will be available for future
computations, so it is playing an output role.

The substitution of variables that are in syntactic constructs of type
pattern (i.e., in a basic graph pattern) has another issue: A variable
\texttt{?x} that is replaced by a value in a basic graph pattern
\texttt{P} does not appear in the solutions of evaluating
\texttt{P}. Thus, the domain of \texttt{P} will be reduced after the
substitution.

This, reduction in the domain of a graph pattern after a substitution,
may produce odd results. Indeed, let \texttt{P} and \texttt{Q} be
respectively the basic graph patterns \texttt{\string{?x :p
    ?y\string}} and \texttt{\string{?y :p ?z\string}}. Let
$\mathtt{P}'$ and $\mathtt{Q}'$ be the results of substituting
\texttt{?y} by \texttt{:b} in \texttt{P} and \texttt{Q}, respectively.
Let $\mu$ be the solution $\{\text{\tt?y}\mapsto\text{\tt:b}\}$. Then,
the graph pattern \texttt{P.Q} has less solutions than
\texttt{P$'$.Q$'$} over the dataset
$\{(\text{\tt:a},\text{\tt:p},\text{\tt:b}),
(\text{\tt:b},\text{\tt:p},\text{\tt:c})\}$. This contradicts, the
intuition that substituting variables with values restrict the results.

\subsection{Normalization}

The normalization of a graph pattern or expression \texttt{P} is
defined to avoid variables with role ambiguity (i.e., that has
simultaneously input and output roles) by changing the structure of
\texttt{P} and replacing every variable occurring in \texttt{P} with a
different fresh variable for each scope that can be determined for the
variable. After the normalization process, variables that can be
substituted will occur only in syntactical constructs of type
expression, so solving the issue described at the end of the previous
section.

\begin{definition}[Normalization]
The normalization of the pattern $\mathtt{P}$, that we denote as
$\mathrm{norm}(\mathtt{P})$, is a 
 triple $(\mathtt{P}',d,g)$, where $\mathtt{P}'$ is a pattern whose  variables must be all
fresh and $d$ and $g$ are partial functions whose domain and ranges
are as follows:
\begin{align*}
  d &:\mathrm{var}(\mathtt{P}')\rightarrow\mathrm{dom}(\mathtt{P}),\\
  g &:\mathrm{var}(\mathtt{P}')\rightarrow\mathrm{var}(\mathtt{P}),
\end{align*}
$d$ is surjective and the domains of $d$ and $g$ are disjoint.
\end{definition}

Intuitively, $d$ and $g$ are functions that associate (record) the
correspondence of the fresh
variables of $\mathtt{P}'$ with the corresponding original variables
$\mathtt{P}$. The function $d$ represents occurrences of variables
that are in the solutions of \texttt{P} and $g$
represents occurrences of variables that
can be substituted by values that $\mu$ maps. The sets
$\mathrm{range}(d)$ and $\mathrm{range}(g)$ could have elements in
common. For example, if \texttt{P} is the graph pattern \texttt{Q.R}
then a variable can be in the domain of \texttt{Q} and
simultaneously be a global variable in \texttt{R}, 

To give an intuition, here there is an illustration of a 
normalization in a simple case. Let \texttt{P} be:
\begin{center}
  \texttt{\string{:a :p ?x\string} $.$ \string{:b :q ?y FILTER (?y < ?x)\string}}.
\end{center}
The result of normalizing \texttt{P} is $(\mathtt{P}',d,g)$
where $\mathtt{P}'$ is
\begin{center}
  \texttt{\string{:a :p $x_1$\string} $.$ \string{:b :q $y_1$ FILTER ($y_1$ < $x_2$)\string}}
\end{center}
and $d$ and $g$ are respectively the functions
\begin{align*}
  d&:=\{x_1\mapsto\text{\tt?x},\, y_1\mapsto \text{\tt?y}\},\\
  g&:=\{x_2\mapsto\text{\tt?x}\}.
\end{align*}
(We use a different notation for variables $\mathtt{P}'$ to stress
the idea that they are fresh.) Note that in the pattern $\mathtt{P}'$ in  $(\mathtt{P}',d,g)$
each variable plays a unique role, and the functions $d,g$ ``tell''
what is the role of each variable and their relationships.

Note that this example uses a particular normalization according with
a specific semantics. An alternative semantics may produce a different
normalization (which is only designed to make  the role of each
variable independent of its occurrence).

\subsection{Substitution and correlated evaluation}

We need a pair of notations before introducing the main notions.
Given a partial function $f$ that maps variables to variables and an
structure of expression $A$ where some of this variables occur, then
$f(A)$ denotes the result of renaming consistently in $A$ every
variable $x\in\mathrm{dom}(f)$ by $f(x)$.  Functions can be viewed a
set of ordered pairs.  We will use the notation $x\mapsto f(x)$
instead of $(x,f(x))$ to stress the notion of mapping. Thus, the
symbol $\varnothing$ (used commonly to denote empty sets) also denotes
empty functions.

\medskip\noindent
Now we are ready to present our main notion:

\begin{definition}[Mapping substitution]
  Let \texttt{P} be a graph pattern, $\mu$ a solution mapping
  and 
  $d$ and $g$ be functions that map variables to variables. Then
  $\mu(\mathtt{P},d,g)$ is the graph pattern $d(\mathtt{P}')$, where
  $\mathtt{P}'$ is the graph pattern resulting of the following
  substitutions in $\mathtt{P}$:
  \begin{enumerate}
  \item For each binding $x\mapsto\texttt{?x}$ in $g$ substitute every
    occurrence of \texttt{bound($x$)} by \texttt{TRUE} if
    $\text{\tt?x}\in\mathrm{dom}(\mu)$ or by \texttt{FALSE} if
    $\text{\tt?x}\not\in\mathrm{dom}(\mu)$.
  \item Then, for each binding $x\mapsto\texttt{?x}$ in $g$ substitute
    every occurrence of $x$ by $\mu(\text{\tt?x})$ if
    $\text{\tt?x}\in\mathrm{dom}(\mu)$ or by \texttt{?x} if
    $\text{\tt?x}\not\in\mathrm{dom}(\mu)$.
  \end{enumerate}
\end{definition}

\begin{definition}[Main: Correlated graph pattern or expression]
  \label{main:defn}
  Let \texttt{P} be a graph pattern or expression, $\mu$ be a solution and
  $\mathrm{norm}$ be a function that receives a graph pattern and
  returns triple $(\mathtt{P}',d,g)$ where $\mathtt{P}'$ is a graph
  pattern and $d$ and $g$ are functions that map variables to
  variables. Then:
\begin{align*}
\bind(\mathtt{P},\mu)=
\left\{
  \begin{array}{ll}
    \mu(\norm(\mathtt{P}))\,.\,\mu|_{\mathrm{dom}(\mathtt{P})} &
    \text{if }\mathtt{P}\text{ if a graph pattern}, \\
    \mu(\norm(\mathtt{P})) &
    \text{if }\mathtt{P}\text{ if an expression}.
  \end{array}
\right.
\end{align*}
\end{definition}

\bigskip\noindent
Note that $\mu|_{\mathrm{dom}(\mathtt{P})}$ denotes the
inline data that codify exactly the multiset containing 
the solution $\mu|_{\mathrm{dom}(\mathtt{P})}$ with multiplicity
1. For example, if $\mu|_{\mathrm{dom}(\mathtt{P})}$ is the solution
$\{\text{\tt?x}\mapsto1,\text{\tt?y}\mapsto2\}$ then it is codified as
\texttt{VALUES (?x ?y) \string{(1 2)\string}}.

The function $\bind$ can be used in any place where the function
$\mathrm{substitute}$ is used by the Standard Spec. For example, given a
dataset $D$, the graph patterns \texttt{P} and \texttt{Q} and the
expression \texttt{E}, then:
\begin{align*}
\semantics{\text{\tt P FILTER (EXISTS \string{Q\string})}}_D
&=\{
\mu \in \semantics{\mathtt{P}}_D
\mid
\semantics{\bind(\mathtt{Q},\mu)}_D \text{ is not empty}
\}\\
\semantics{\text{\tt P BIND (E AS ?x)}}_D
&= \semantics{P}_D \Join \{\{\text{\tt?x}\mapsto\semantics{\bind(\mathtt{E},\mu)}_D\}\}
\end{align*}

In what follows, we present three variants of the normalization
function. Each one, according to Definition \ref{main:defn} will give
a particular semantics.
Given a graph pattern \texttt{P} and its normalization
$(\mathtt{P}',d,g)$, these variants differ essentially in the variables
occurring in \texttt{P} that are included in the range of
the function $g$. Intuitively, variables that are excluded of the
ranges of $d$ and $g$ can be considered local, because the
normalization renames them to fresh variables and does not record the
original names.

\subsection{Semantics S1}
\label{sec:s1}

According S1 all variables that are not in the domain of a graph
pattern are considered local.  Thus, the normalization in S1 is
defined as follows:

\begin{definition}[Normalization in S1]
  Given a graph pattern \texttt{P}, then $\mathrm{norm}(\mathtt{P})$ is
  $(\mathtt{P}',d,\varnothing)$ where:
  \begin{enumerate}
  \item $d$ is a surjective function that maps fresh variables to
    variables in $\mathrm{dom}(\mathtt{P})$.
  \item $\mathtt{P}'$ is $h(d^{-1}(\mathtt{P}))$ where $h$ is a function
    that maps variables in
    $\mathrm{var}(\mathtt{P})\setminus\mathrm{dom}(\mathtt{P})$ to fresh
    variables.
  \end{enumerate}
\end{definition}

At the end of this procedure is ensure that all local variables in
\texttt{P} are substituted $\mathtt{P}'$ with fresh variables that
will be not substituted again because the third component of the
normalization is empty. This is summarized in the following result.

\begin{lemma}
  According the semantics S1, given
  a graph pattern \texttt{P}, a solution mapping $\mu$ and a dataset $D$, then:
  \[
  \semantics{\bind(\mathtt{P},\mu)}_D=\semantics{\mathtt{P}}_D\Join\{\mu_{|\mathrm{dom}(\mathtt{P})}\}
  \]
\end{lemma}

\subsection{Semantics S2}
\label{sec:s2}

Before defining this semantics we need some definitions that will help
us in the notation.

\begin{definition}[The filter clause]
  Given two variables \texttt{?y} and \texttt{?y} then
  $F_{\text{\tt?x?y}}$ is the operator
  \texttt{FILTER (!(bound(?x) \&\& bound(?y)) || ?x = ?y)}.
\end{definition}

The operator $F_{\text{\tt?x?y}}$ help us to rewrite a variable that
is in the domain of a graph pattern as a variable whose visibility is
global according S2 and S3. For example, let \texttt{P} and \texttt{Q}
be the graph patterns \texttt{\{a :p ?x\}} and
\texttt{\{:x :p ?y\} $F_{\tt?x?y}$}, respectively. Then, intuitively
\(\semantics{\mathtt{P}}_\mu=\semantics{\mathtt{Q}}_\mu\) for every
mapping $\mu$.

\begin{definition}[Consequently renaming]
  Let be $f$ and $g$ be two functions that map variables to
  variables where $g$ is injective. Then, $\mathrm{cr}(f,g)$ is the function
  $g^{-1}|_A\cdot f|_A$ where $A$ is
  $\mathrm{range}(f)\cap\mathrm{range}(g)$
  and ``$\cdot$'' denotes the composition of functions\footnote{
    Note that that is if $x\in\mathrm{dom}(f)$ and $f(x)\in\dom(g)$
    then $(f \cdot g)(x)=g(f(x))$.
  }.
\end{definition}

If a graph pattern \texttt{P} is composed of the graph patterns
\texttt{Q} and \texttt{R}, then results natural defining the
normalization of \texttt{P} as a composition of the respective
normalizations $\mathtt{Q}'$ and $\mathtt{R}'$ of its
components. Because the normalization of these components are
performed independently, the variables in the domain may be different,
thus it is needed to rename variables in one of the components to make
both renaming consequent in the outputs of both components. The
following lemma show that given to renamings $f$ and $g$ where $g$ is
injective, then $\mathrm{cr}(f,g)$ can be used to generate a function
$g'$ that is compatible with $f$, that is $(\mathrm{cr}(f,g))(g)$.

\begin{lemma}
  Given two functions $f$ and $g$ that map variables to variables
  where $g$ is injective,
  then $(\mathrm{cr}(f,g))(g|_A) = f|_A$.
\end{lemma}

At this point we are ready to proceed with the formalization of the
normalization in the semantics S2.

\begin{definition}[Normalization in S2]
Given a graph pattern or expression \texttt{P}, then:
\begin{enumerate}
\item If \texttt{P} is a basic graph pattern, then
  $\mathrm{norm}(\mathtt{P})$ is $(d^{-1}(\mathtt{P}),d,\varnothing)$, where
  $d$ is a function that contains a binding $x\mapsto\text{\tt?x}$ for
  every variable \texttt{?x} in $\mathrm{var}(\mathtt{P})$.
\item If \texttt{P} is \texttt{SELECT X WHERE \string{Q\string}}
  (where \texttt{X} is a list of variables), then
  \(\mathrm{norm}(\mathtt{P})\) is \((\mathtt{P}',d_{\mathtt{P}},g_{\mathtt{P}})\),
  where:
  \begin{align*}
    \mathtt{P}' &= \text{\tt SELECT X$'$ WHERE \string{Q$'$\string}}\\
    (\mathtt{Q}',d_{\mathtt{Q}},g_{\mathtt{Q}}) &= \mathrm{norm}(\mathtt{Q})\\
    d_{\mathtt{P}} &= d_{\mathtt{Q}}|_{d_{\mathtt{Q}}^{-1}(\mathrm{dom}(\mathtt{P}))}\\
    g_{\mathtt{P}} &= g_{\mathtt{Q}}\\
    \mathtt{X}' &= d_{\mathtt{P}}^{-1}(\mathtt{X})
  \end{align*}
  Note that $d_{\mathtt{Q}}^{-1}(\mathrm{dom}(\mathtt{P}))$ is the
  preimage in $d_{\mathtt{Q}}$ of $\mathrm{dom}(\mathtt{P})$. That is,
  the set of variables used in \texttt{Q} to rename variables that are
  projected in the solution of \texttt{P}. Thus,
  $d_{\mathtt{Q}|d_{\mathtt{Q}}^{-1}(\mathrm{dom}(\mathtt{P}))}$ is
  the renaming of variables used in \texttt{Q}, restricted to the
  domain of \texttt{P}. Similarly, $d_{\mathtt{P}}^{-1}(\mathtt{X})$
  is renaming of variables in \texttt{X} that is consecuent with the
  renaming done in \texttt{Q}.

  Bindings $x\mapsto\text{\tt?x}$ in $d_{\mathtt{Q}}$ such that
  \texttt{?x} is not in the domain of \texttt{P} are not included
  in $d_{\mathtt{P}}$ nor in $g_{\mathtt{Q}}$. This, is interpreted as
  that the cocurrences of \texttt{?x} associated to this bindings are
  assumed local.

\item If \texttt{P} is $\mathtt{Q}\circ\mathtt{R}$ where $\circ$ is
  `\texttt{.}', \texttt{OPTIONAL} or \texttt{UNION}, then
  \(\mathrm{norm}(\mathtt{P})\) is
  \((\mathtt{P}',d_{\mathtt{P}},g_{\mathtt{P}})\), where:
  \begin{align*}
    \mathtt{P}' &= \mathtt{Q}' \circ f(\mathtt{R}')\\
    (\mathtt{Q}',d_{\mathtt{Q}},g_{\mathtt{Q}}) &= \mathrm{norm}(\mathtt{Q})\\
    (\mathtt{R}',d_{\mathtt{R}},g_{\mathtt{R}}) &= \mathrm{norm}(\mathtt{R})\\
    d_{\mathtt{P}} &= d_{\mathtt{Q}}\cup f(d_{\mathtt{R}})\\    
    g_{\mathtt{P}} &= g_{\mathtt{Q}}\cup g_{\mathtt{R}}\\
    f &= \mathrm{cr}(d_{\mathtt{Q}},d_{\mathtt{R}})
  \end{align*}
  Note that $f$ is a renaming that ensure that the
  normalizations of \texttt{Q} and \texttt{R} use the same common
  domain variables when they are combained. 
\item If \texttt{P} is \texttt{Q MINUS R}, then
  \(\mathrm{norm}(\mathtt{P})\) is
  \((\mathtt{Q'\;MINUS}\;f(\mathtt{R}'),d_{\mathtt{P}},g_{\mathtt{P}})\), where:
  \begin{align*}
    (\mathtt{Q}',d_{\mathtt{Q}},g_{\mathtt{Q}}) &= \mathrm{norm}(\mathtt{Q})\\
    (\mathtt{R}',d_{\mathtt{R}},g_{\mathtt{R}}) &= \mathrm{norm}(\mathtt{R})\\
    d_{\mathtt{P}} &= d_{\mathtt{Q}}\\    
    g_{\mathtt{P}} &= g_{\mathtt{Q}}\cup g_{\mathtt{R}}\\
    f &= \mathrm{cr}(d_{\mathtt{Q}},d_{\mathtt{R}}).
  \end{align*}
  The function $f$ is a renaming that ensures that the variables used
  in the variables that are common in domains of the normalizations of
  \texttt{Q} and \texttt{R} are renamed to the the same fresh
  variables.

  Note that variables that are in
  $\mathrm{dom}(\mathtt{R})\setminus\mathrm{dom}(\mathtt{Q})$ are
  replaced with fresh variables that are not included in the domains
  of $d_{\mathtt{P}}$ and $g_{\mathtt{P}}$. Thus, they are assumed
  local.
\item If \texttt{P} is \texttt{GRAPH u \{Q\}} where \texttt{u} is an
  IRI, then $\mathrm{norm}(\mathtt{P})$ is
  \((\text{\tt GRAPH u \string{Q$'$\string}},d,g)\),
  where $\mathrm{norm}(\mathtt{Q})=(\mathtt{Q}',d,g)$.
\item If \texttt{P} is \texttt{GRAPH ?x \{Q\}},
  then $\mathrm{norm}(\mathtt{P})$ is
  \((\text{\tt GRAPH $x$ \string{Q$'$\string}},d_{\mathtt{P}},g_{\mathtt{P}})\),
  where:
  \begin{align*}
    (\mathtt{Q}',d_{\mathtt{Q}},g_{\mathtt{Q}}) &= \mathrm{norm}(\mathtt{Q})\\
    d_{\mathtt{P}} &=
    \left\{
      \begin{array}{ll}
        d_{\mathtt{Q}} & \text{if \tt?x}\in\mathrm{range}(d)\\
        d_{\mathtt{Q}}\cup\{x\mapsto\text{\tt?x}\} & \text{otherwise ($x$ is fresh)}
      \end{array}
    \right.\\    
    g_{\mathtt{P}} &= g_{\mathtt{Q}}.
  \end{align*}
\item If \texttt{P} is \texttt{SERVICE u \{Q\}} where \texttt{u} is an
  IRI, then $\mathrm{norm}(\mathtt{P})$ is
  \((\mathtt{P}',d,g)\),
  where $\mathrm{norm}(\mathtt{Q})=(\mathtt{Q}',d,g)$
  and \(\mathtt{P}'=\text{\tt SERVICE u \string{Q$'$\string}}\).
\item If \texttt{P} is \texttt{Q FILTER (C)}, then
  $\mathrm{norm}(\mathtt{P})$ is \((\mathtt{P}',d_{\mathtt{P}},g_{\mathtt{P}})\)
  where:
  \begin{align*}
    \mathrm{P}' &= \text{\tt Q$'$ FILTER ($f(\mathtt{C}')$)},\\
    (\mathtt{Q}',d_{\mathtt{Q}},g_{\mathtt{Q}}) &= \mathrm{norm}(\mathtt{Q}),\\
    (\mathtt{C}',\varnothing,g_{\mathtt{C}}) &= \mathrm{norm}(\mathtt{C}),\\
    d_{\mathtt{P}} &= d_{\mathtt{Q}},\\
    g_{\mathtt{P}} &= g_{\mathtt{Q}}\cup f(g_{\mathtt{C}}),\\
    f &= \mathrm{cr}(d_{\mathtt{Q}},d_{\mathtt{C}}).
  \end{align*}
\item If \texttt{P} is \texttt{VALUES (X) \{B\}} where \texttt{X} is a
  list of variables and \texttt{B} is a list of bindings to the
  variables, then $\mathrm{norm}(\mathtt{P})$ is
  $(\mathtt{P}',d,\varnothing)$, where $d$ has a binding
  $x\mapsto\text{\tt?x}$ for each variable \texttt{?x} in \texttt{X} and
  $\mathtt{P'}=d^{-1}(\mathtt{P})$.
\item If \texttt{P} is \texttt{Q BIND (E AS ?x)} then
  $\mathrm{norm}(\mathtt{P})$ is
  \((\mathtt{P}',d_{\mathtt{P}},g_{\mathtt{P}})\) where:
  \begin{align*}
    \mathrm{P}' &= \text{\tt Q$'$ BIND ($f(\mathtt{E}')$ AS $x$)}\\
    (\mathtt{Q}',d_{\mathtt{Q}},g_{\mathtt{Q}}) &= \mathrm{norm}(\mathtt{Q})\\
    (\mathtt{E}',\varnothing,g_{\mathtt{E}}) &= \mathrm{norm}(\mathtt{E})\\
    d_{\mathtt{P}} &= d_{\mathtt{Q}}\cup\{x\mapsto\text{\tt?x}\},\\
    g_{\mathtt{P}} &= g_{\mathtt{Q}}\cup f(g_{\mathtt{E}}),\\
    f &= \mathrm{cr}(d_{\mathtt{Q}},d_{\mathtt{E}}).
  \end{align*}
\item If \texttt{P} is an expression and
  $\{\mathtt{Q}_1,\dots\mathtt{Q}_n\}$ is the set of graph patterns
  that are directely contained into maximal occurrences of \texttt{EXITS} clauses in
  \texttt{P} (we say that an \texttt{EXITS} clause occurrence $i$ is maximal in
  \texttt{P} if does not occur another \texttt{EXITS} clause $j$ containing $i$
  in \texttt{P}). Then $\mathrm{norm}(\mathtt{P})$ is
  $(\mathtt{P}_n,\varnothing,g_n)$ computed recursively as follows:
  \begin{enumerate}
  \item Let $\mathtt{P}_0$ be \texttt{P} and $g_0$ be the function
    that include a binding $x\mapsto\text{\tt?x}$ for every variable
    \texttt{?x} in \texttt{P} that does not occur in any of the graph
    patterns $\{\mathtt{Q}_1,\dots\mathtt{Q}_n\}$.
  \item For each $\mathtt{Q}_k$ of the graph patterns in the maximal
    \texttt{EXISTS} clauses, let $(\mathtt{Q}'_k,d'_k,g'_k)$ be
    $\mathrm{norm}(\mathtt{Q}_k)$. Then, let $\mathtt{P}_k$ the result
    of replacing in $\mathtt{P}_{k-1}$ the occurrence of
    $\mathtt{Q}_k$ by $f(\mathtt{Q}_k)\,F_{x_1y_1}\dots F_{x_my_m}$
    where:
    \begin{align*}
      f &= \mathrm{cr}(g_0,g'_k),\\
      g_k &= g_{k-1}\cup f(g'_k) \cup h(d'_k)
    \end{align*}    
    and $h$ be a function $\{y_1\mapsto x_1,\dots,y_m\mapsto x_m\}$ that
    has a binding $y\mapsto x$ for each binding $x\mapsto\text{\tt?x}$
    in $d'_k$.
  \end{enumerate}
\end{enumerate}
\end{definition}

\subsection{Semantics S3}
\label{sec:s3}

In the rules 2 and 4, the semantics S2 assumes that pattern
occurrences that are not in the domain of a graph pattern are
local, so they are not included in the global bindings. On the
contrary, S3 assumes that they are global, so the query is modified to
move variables from graph occurrences to expression occurrences using
operations $F_{xy}$. After this transformation, substitution can be
applied in the same way that in semantics S2.

\begin{definition}[Normalization in S3]
Given a graph pattern or expression \texttt{P}, then the normalization
of \texttt{P} is computed using the same rules enumerated in the
definition of the normalization of S3, except the rules 2 and 4, that
are replaced with the following rules:
\begin{enumerate}
\item[2.] If \texttt{P} is \texttt{SELECT X WHERE \string{Q\string}}
  (where \texttt{X} is a list of variables), then
  \(\mathrm{norm}(\mathtt{P})\) is \((\mathtt{P}',d_{\mathtt{P}},g_{\mathtt{P}})\),
  where:
  \begin{align*}
    \mathtt{P}' &= \text{\tt SELECT X$'$ WHERE \string{Q$'\,F_{x_1y_1}\dots F_{x_my_m}$\string}}\\
    (\mathtt{Q}',d_{\mathtt{Q}},g_{\mathtt{Q}}) &= \mathrm{norm}(\mathtt{Q})\\
    d_{\mathtt{P}} &= d_{\mathtt{Q}|\mathrm{dom}(\mathtt{P})}\\
    g_{\mathtt{P}} &= g_{\mathtt{Q}}\cup h(d_{\mathtt{Q}|\mathrm{dom}(\mathtt{Q})\setminus\mathrm{dom}(\mathtt{P})})\\
    \mathtt{X}' &= d_{\mathtt{P}}^{-1}(\mathtt{X})
  \end{align*}
  and $h$ is a function $\{y_1\mapsto x_1,\dots,y_m\mapsto x_m\}$ that
  has a binding $y\mapsto x$ for each binding $x\mapsto\text{\tt?x}$
  in $d_{\mathtt{Q}|\mathrm{dom}(\mathtt{Q})\setminus\mathrm{dom}(\mathtt{P})}$.

\item[4.] If \texttt{P} is \texttt{Q MINUS R}, then
  $\mathrm{norm}(\mathtt{P})$ is
  \((\mathtt{P}',d_{\mathtt{P}},g_{\mathtt{P}})\) where:
  \begin{align*}
    \mathtt{P}' &= \text{\tt Q$'$ MINUS \string{$f(\mathtt{R}')$
        $F_{x_1y_1} \dots F_{x_my_m}$)\string}} \\
    (\mathtt{Q}',d_{\mathtt{Q}},g_{\mathtt{Q}}) &=
    \mathrm{norm}(\mathtt{Q}) \\
    (\mathtt{R}',d_{\mathtt{R}},g_{\mathtt{R}}) &=
    \mathrm{norm}(\mathtt{R})\\
    d_{\mathtt{P}} &= d_{\mathtt{Q}}\\
    g_{\mathtt{P}} &= g_{\mathtt{Q}}\cup g_{\mathtt{R}} \cup h(d_{\mathtt{R}|\mathrm{dom}(\mathtt{R})\setminus\mathrm{dom}(\mathtt{P})})\\
    f &= \mathrm{cr}(d_{\mathtt{Q}},d_{\mathtt{R}})
  \end{align*}
  and $h$ is a function $\{y_1\mapsto x_1,\dots,y_m\mapsto x_m\}$ that
  has a binding $y\mapsto x$ for each binding $x\mapsto\text{\tt?x}$
  in $d_{\mathtt{R}|\mathrm{dom}(\mathtt{R})\setminus\mathrm{dom}(\mathtt{P})}$.
\end{enumerate}
\end{definition}

\section{Substitution and blank nodes}
\label{sec:blank-nodes}

Peter F. Patel-Schneider noticed in the W3C mailing list of SPARQL
that substitution has problems with blank nodes and a semantics where
every variable can be substituted. Let \texttt{P} be the inner graph
pattern of a query and \texttt{?x} be a variable that can be
substituted in \texttt{P}, in particular with a blank node
\texttt{\_:b}. Then, there are the following options:
\begin{enumerate}
\item \texttt{?x} occurs in a basic graph pattern and is
  substituted by \texttt{\_:b}. Then, according the specification
  \texttt{\_:b} is interpreted as an existential variable and scoped
  to the basic graph pattern. Thus, \texttt{\_:b} may represent any
  element on the graph, not only \texttt{\_:b}.
\item \texttt{?x} is in an expression occurrence. Then, the
  substitution of \texttt{?x} by \texttt{\_:b} restricts the resulting
  bindings to whose where \texttt{?x} is bound to \texttt{\_:b}.
\item \texttt{?x} is in the domain of \texttt{P}. Then, results of
  \texttt{P} that are not compatible with
  $\{\text{\tt?x}\mapsto\text{\tt\_:b}\}$ are discarded.
\end{enumerate}
The first case results contradictory with the other two. In fact, it
does not restrict the variable \texttt{?x} as the other do (and as it
is expected for substitution).

In the semantics S3 proposed in this technical report a variable
\texttt{?x} occurring in a pattern occurrence is considered
replaceable with values that come from the current solution
mapping. However, before this substitution the normalization process
moves \texttt{?x} from the pattern occurrence to an expression
occurrence using a renaming of \texttt{?x} to \texttt{?y} and then
using an operator $F_{\text{\tt?x?y}}$. Thus, we can conclude the
following lemma:

\begin{lemma}
  The semantics S1, S2 and S3 are safe respect with the blank nodes
  substitution issue.
\end{lemma}

\section{Correlation in implementations}
\label{sec:examples}

This section presents examples of how the proposed semantics work and
how different implementations match them.
Queries presented in this section are run against the RDF
graph depicted in Figure \ref{fig:rdf-graph}.

For each query, the actual results given by each implementation is
shown at the end of the section.

\subsection*{Example 1}
\begin{Verbatim}[fontsize=\small]
SELECT ?parent
WHERE { ?parent :country :j
        FILTER ( EXISTS { ?child :parent ?parent })}
\end{Verbatim}
This query gets people of country \texttt{:j} having children. That
is, select people that has solutions for the inner query. The variable
\texttt{?people} is in-domain in the inner graph pattern for all
semantics. Thus, the results of inner graph pattern are filtered to be
compatible with solutions of the outer query $\mu_a$ and $\mu_b$. The
only solution that has results for the inner graph pattern is $\mu_b$
in each of the three semantics S1, S2 and S3.

\subsection*{Example 2}
\begin{Verbatim}[fontsize=\small]
SELECT ?parent
WHERE { ?parent :country :j
        FILTER ( EXISTS { SELECT ?child
                          WHERE { ?child :parent ?parent }})}
\end{Verbatim}
This query is similar to the presented in Example~1. However, in this
case the variable \texttt{?parent} in the inner query is in local
according with S1 and S2, and global according with S3. Thus, the
solutions are $\{\mu_a,\mu_b\}$ for S1 and S2.  On the other hand, the
result is $\{\mu_b\}$ for S3.

\subsection*{Example 3}
\begin{Verbatim}[fontsize=\small]
SELECT ?parent
WHERE { ?parent :country :j
        FILTER ( EXISTS { SELECT ?child
                          WHERE { ?child :parent ?chparent
                                  FILTER (?chparent = ?parent) }})}
\end{Verbatim}
This query is similar to the presented in the previous
examples. However, in this case the variable \texttt{?parent} in the
inner query is global according with S2 and S3, and local according
with S1. Thus, the solutions are $\{\mu_b\}$ for S2 and S3. On the
other hand, the result is $\{\}$ for S1. Indeed, the filter condition
\texttt{?chparent = ?parent} gets an error, because \texttt{?parent}
is unbound .

\subsection*{Example 4}
\begin{Verbatim}[fontsize=\small]
SELECT ?parent
WHERE { ?parent :country :j
        FILTER ( EXISTS { SELECT ?child
                          WHERE { ?child :parent ?chparent
                                  FILTER (bound(?parent)) }})}
\end{Verbatim}
This query checks if the variable \texttt{?parent} is bound in the
inner sub-select with the build-in function
\texttt{bound(?parent)}. Thus, this query returns $\{\mu_a,\mu_b\}$ in
S2 and S3 and returns $\{\}$ in S1.

\section*{Example 5}
\begin{Verbatim}[fontsize=\small]
SELECT ?parent
WHERE { ?parent :country :j
        FILTER ( EXISTS { SELECT ?child
                          WHERE { ?child :parent ?chparent
                                  FILTER (?chparent = ?parent &&
                                          bound(?parent)) }})}
\end{Verbatim}
This query is equivalent to the query of Example 3 according the
studied semantics. The addition of the condition
\texttt{bound(?parent)} must not affect the result.

\subsection*{Example 6}
\begin{Verbatim}[fontsize=\small]
SELECT ?parent
WHERE { ?parent :country :j
        FILTER ( EXISTS { SELECT ?child ?chparent
                          WHERE { ?child :parent ?chparent
                                  FILTER (?parent = 1 ||
                                          ?parent != 1 )}})}
\end{Verbatim}
In this example, the filter clause \texttt{?parent = 1 || ?parent != 1}
is a tautology when \texttt{?parent} is bound. Otherwise, both
sides of the disjunction are evaluated as error so the whole clause
gets an error. Thus, the output of this query is $\{\mu_a,\mu_b\}$
according S2 and S3 and $\{\}$ according S1.

\subsection*{Example 7}
\begin{Verbatim}[fontsize=\small]
SELECT ?parent
WHERE { ?parent :country :j
        FILTER ( EXISTS { SELECT *
                          WHERE { ?child :parent ?chparent
                                  FILTER (?parent = 1 ||
                                          ?parent != 1 )}})}
\end{Verbatim}
This example is equivalent with Example 6. In fact, the semantics of
the wildcard `\text{*}' is the list of all variables that are
in-domain of the query.

\subsection*{Example 8}
\begin{Verbatim}[fontsize=\small]
SELECT ?parent
WHERE { ?parent :country :j
        FILTER ( EXISTS { SELECT ?child
                          WHERE { ?child :parent ?parent
                                  FILTER (?parent = :c)}})}
\end{Verbatim}
In this query the variable \texttt{?parent} has three occurrences. The
first is in the outer graph pattern and the other two in the inner
graph pattern. In any of the semantics \texttt{?parent} is bound to
\texttt{:a} and \texttt{:b} in the solutions of the outer graph
pattern.

According S2 and S3 the variable \texttt{?parent} is local in the
inner graph pattern. Thus it is bound to \texttt{:a}, \texttt{:b} and
\texttt{:c} in the inner graph pattern. Then, the filter clause of the
inner graph pattern is true for \texttt{:c}. Thus, the result of this
query is $\{\mu_a,\mu_b\}$.

On the other hand, according S3 the variable \texttt{?parent} is
global in the inner graph pattern. So, it is replaced with the values
comming from the outer graph pattern. None of this values satisfy the
condition \texttt{?parent = :c}. Thus, the result of this query is
$\{\}$.

\subsection*{Example 9}
\begin{Verbatim}[fontsize=\small]
SELECT ?parent
WHERE { ?parent :country :j
        FILTER ( EXISTS { SELECT ?child
                          WHERE { ?child :parent ?parent
                                  FILTER (EXISTS{?parent :parent :d})}})}
\end{Verbatim}
On the dataset used in these examples, This query seems to be
equivalent to the previous query (presented in Example 8), because the
graph pattern \texttt{?parent :parent :d} only has solutions if
\texttt{?parent} is \texttt{:c}.  Thus, in S2 and S3 the result of
this query is $\{\}$. On the other hand, in S1 the result of this
query is $\{\mu_a,\mu_b\}$.

\subsection*{Example 10}
\begin{Verbatim}[fontsize=\small]
SELECT *
WHERE { { { ?x :p ?y } OPTIONAL { ?y :q ?z } }
        FILTER ( EXISTS { ?z :r ?v } ) }
\end{Verbatim}
Consider the following RDF graph that is the dataset $D$ where we will
evaluate this query.

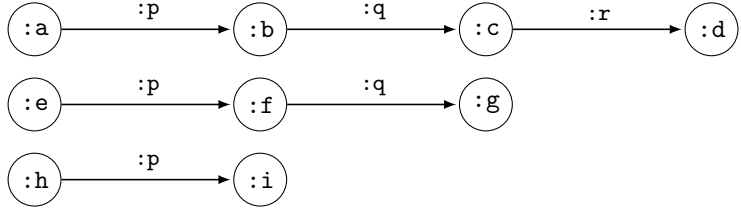
\begin{figure}[h!]
\begin{center}
\begin{tikzpicture}
\begin{scope}[auto, every node/.style={draw,circle,minimum size=2em,inner sep=1},node distance=2cm]
\def\margin {8}
 \node[draw,circle](a) at (0,0) {\texttt{:a}};
 \node[draw,circle](b) at (3,0) {\texttt{:b}};
 \node[draw,circle](c) at (6,0) {\texttt{:c}};
 \node[draw,circle](d) at (9,0) {\texttt{:d}};
 \node[draw,circle](e) at (0,-1) {\texttt{:e}};
 \node[draw,circle](f) at (3,-1) {\texttt{:f}};
 \node[draw,circle](g) at (6,-1) {\texttt{:g}};
 \node[draw,circle](h) at (0,-2) {\texttt{:h}};
 \node[draw,circle](i) at (3,-2) {\texttt{:i}};
\end{scope}

\begin{scope}[->,>=latex,semithick,shorten >=1pt, auto]
\draw(a) to node {{\small\texttt{:p}}} (b);
\draw(b) to node {{\small\texttt{:q}}} (c);
\draw(c) to node {{\small\texttt{:r}}} (d);
\draw(e) to node {{\small\texttt{:p}}} (f);
\draw(f) to node {{\small\texttt{:q}}} (g);
\draw(h) to node {{\small\texttt{:p}}} (i);
\end{scope}
\end{tikzpicture}
\end{center}
\caption{RDF graph.}
\end{figure}
The normalization of the inner graph pattern gives the same result in
the three semantics, because the variables \texttt{?z} and \texttt{?v}
are in the domain of the inner graph pattern. Thus, the evaluation of
this query is the set $\Omega$ defined as:
\[
\Omega=
\{\mu \in \semantics{\mathtt{P}}_D \mid
\semantics{\mathtt{Q}}_D \Join \{\mu\}
\text{ is not empty}\},
\]
where \texttt{P} and \texttt{Q} are the outer and inner graph
patterns, respectively.
$\semantics{\mathtt{P}}_D$ and $\semantics{\mathtt{Q}_D}$ are
respectively the sets
$\{\mu_{\mathtt{abc}},\mu_{\mathtt{efg}},\mu_{\mathtt{hi}}\}$ and
$\{\mu_{\mathtt{cd}}\}$ where:
\begin{align*}
\mu_{\mathtt{abc}} &=
\{\text{\tt?x}\mapsto\text{\tt:a},
\text{\tt?y}\mapsto\text{\tt:b},
\text{\tt?z}\mapsto\text{\tt:c}\},\\
\mu_{\mathtt{efg}} &=
\{\text{\tt?x}\mapsto\text{\tt:e},
\text{\tt?y}\mapsto\text{\tt:f},
\text{\tt?z}\mapsto\text{\tt:g}\},\\
\mu_{\mathtt{hi}} &=
\{\text{\tt?x}\mapsto\text{\tt:h},
\text{\tt?y}\mapsto\text{\tt:i}\},\\
\mu_{\mathtt{cd}} &=
\{\text{\tt?z}\mapsto\text{\tt:c},
\text{\tt?v}\mapsto\text{\tt:d}\}.
\end{align*}
Thus, $\Omega=\{\mu_{\mathtt{abc}},\mu_{\mathtt{hi}}\}$.

\subsection*{Summary}

The following table summarizes the results that the example queries
get for each of the studied semantics and the results that the studied
implementations actually output.

\medskip

\begin{small}
\hspace{-4.2em}
\begin{tabular}{rccccccc}
\toprule
\# &
S1 & S2 & S3 &
rdf4j & Virtuoso & Fuseki & Blazegraph \\
\midrule
1 &
$\{\mu_b\}$ &
$\{\mu_b\}$ &
$\{\mu_b\}$ &
$\{\mu_b\}$ &
$\{\mu_b\}$ &
$\{\mu_b\}$ &
$\{\mu_b\}$ \\
2 &
$\{\mu_a,\mu_b\}$ &
$\{\mu_a,\mu_b\}$ &
$\{\mu_b\}$ &
$\{\mu_b\}$ &
$\{\mu_b\}$ &
$\{\mu_a,\mu_b\}$ &
$\{\mu_a,\mu_b\}$ \\
3 &
$\{\}$ &
$\{\mu_b\}$ &
$\{\mu_b\}$ &
$\{\mu_b\}$ &
$\{\mu_b\}$ &
$\{\}$ &
$\{\}$ \\
4 &
$\{\}$ &
$\{\mu_a,\mu_b\}$ &
$\{\mu_a,\mu_b\}$ &
$\{\mu_a,\mu_b\}$ &
$\{\}$ & 
$\{\}$ &
$\{\}$ \\
5 &
$\{\}$ &
$\{\mu_b\}$ &
$\{\mu_b\}$ &
$\{\mu_b\}$ &
$\{\mu_b\}$ & 
$\{\}$ &
$\{\}$ \\
6 &
$\{\}$ & 
$\{\mu_a,\mu_b\}$ & 
$\{\mu_a,\mu_b\}$ & 
$\{\mu_a,\mu_b\}$ & 
$\{\mu_a,\mu_b\}$ & 
$\{\}$ & 
$\{\}$ \\ 
7 &
$\{\}$ & 
$\{\mu_a,\mu_b\}$ & 
$\{\mu_a,\mu_b\}$ & 
$\{\mu_a,\mu_b\}$ & 
$\{\mu_a,\mu_b\}$ & 
$\{\mu_a,\mu_b\}$ & 
$\{\}$ \\ 
8 &
$\{\mu_a,\mu_b\}$ & 
$\{\mu_a,\mu_b\}$ & 
$\{\}$ & 
$\{\}$ & 
$\{\mu_a,\mu_b\}$ & 
$\{\mu_a,\mu_b\}$ & 
$\{\mu_a,\mu_b\}$ \\ 
9 &
$\{\mu_a,\mu_b\}$ & 
$\{\mu_a,\mu_b\}$ & 
$\{\}$ & 
$\{\}$ & 
$\{\}$ & 
$\{\mu_a,\mu_b\}$ & 
$\{\mu_a,\mu_b\}$ \\ 
10 &
$\{\mu_{\mathtt{abc}},\mu_{\mathtt{hi}}\}$ & 
$\{\mu_{\mathtt{abc}},\mu_{\mathtt{hi}}\}$ & 
$\{\mu_{\mathtt{abc}},\mu_{\mathtt{hi}}\}$ & 
$\{\mu_{\mathtt{abc}},\mu_{\mathtt{hi}}\}$ & 
$\{\mu_{\mathtt{abc}}\}$ & 
$\{\mu_{\mathtt{abc}},\mu_{\mathtt{hi}}\}$ & 
$\{\mu_{\mathtt{abc}},\mu_{\mathtt{hi}}\}$ \\ 
\bottomrule
\end{tabular}
\end{small}

\medskip

We distinguish two groups of implementations. Blazegraph and Fuseki
match S1 and Virtuoso and rdf4j matche S3. However, only Blazegraph
and rdf4j match their respectively semantics in all the examples. 

Fuseki agree with S1 except in the query of Example
7 with . The queries in the examples 6 and 7 are equivalent. However,
in Fuseki the results differ. This seems as a bug of Fuseki.

Virtuoso agree with S3 except in the queries of
examples 4, 8 and 10. The result given in the query 4 seems as a bug,
because it is contradictory that \texttt{?parent} is unbound when it
is bound in query 5. The result given in the query
8 require more study because it is not clear if it is a bug or it is
motivated by a different interpretation of the correlation.
The result given in the query 10 shows that there are an special
treatment with unbound variables.

\end{document}